\documentclass[preprint]{aastex}

\slugcomment{Accepted by ApJ Letters, 4 August 2006}

\newcommand{\skipthis}[1]{}

\newcommand{\Msun}{\mbox{$M_{\sun}$}}
\newcommand{\Lsun}{\mbox{$L_{\sun}$}}
\newcommand{\cc}{\mbox{cm$^{-3}$}}  

\newcommand{\um}{\mbox{$\mu$m}}

\shortauthors{Bourke et al.}
\shorttitle{L1521F-IRS Protostar Discovery}

\begin{document}

\title {The Spitzer c2d Survey of Nearby Dense Cores: II: Discovery of a Low
Luminosity Object in the ``Evolved Starless Core'' L1521F}

\author{
Tyler L.\ Bourke\altaffilmark{1}, 
Philip C.\ Myers\altaffilmark{1}, 
Neal J.\ Evans II\altaffilmark{2}, 
Michael M.\ Dunham\altaffilmark{2}, 
Jens Kauffmann\altaffilmark{3},
Yancy L.\ Shirley\altaffilmark{4},
Antonio Crapsi\altaffilmark{5},
Chadwick H.\ Young\altaffilmark{6},
Tracy L.\ Huard\altaffilmark{1}, 
Timothy Y.\ Brooke\altaffilmark{7},
Nicholas Chapman\altaffilmark{8},
Lucas Cieza\altaffilmark{2},
Chang Won Lee\altaffilmark{9},
Peter Teuben\altaffilmark{8},
Zahed Wahhaj\altaffilmark{10}
}

\altaffiltext{1}{Harvard-Smithsonian Center for Astrophysics, 60 Garden
Street, Cambridge, MA 02138; email tbourke@cfa.harvard.edu}
\altaffiltext{2}{University of Texas at Austin, 1 University Station C1400,
Austin, TX 78712-0259}
\altaffiltext{4}{Steward Observatory, University of Arizona, 933 N. Cherry
Ave., Tucson, AZ 85721}
\altaffiltext{3}{Max-Planck-Institut f\"ur Radioastronomie, Auf den H\"ugel
71, D-53121 Bonn, Germany}
\altaffiltext{5}{Leiden Observatory, Postbus 9513, 2300 RA Leiden,
Netherlands}
\altaffiltext{6}{Nicholls State University, P.O. Box 2022, Thibodaux, LA
70310}
\altaffiltext{7}{Division of Physics, Mathematics, and Astronomy, MS
105-24, California Institute of Technology, Pasadena, CA 91125}
\altaffiltext{8}{Department of Astronomy, University of Maryland, College
Park, MD, 20742}
\altaffiltext{9}{Korea Astronomy \& Space Science Institute 61-1
Hwaam-dong, Yusung-gu, Daejeon 305-348, Korea}
\altaffiltext{10}{Department of Physics and Astronomy, Northern Arizona
University, Box 6010, Flagstaff, AZ 86011-6010}

\begin{abstract}
We present Spitzer Space Telescope observations of the ``evolved starless
core'' L1521F which reveal the presence of a very low luminosity object
($L <$0.07 \Lsun).  The object, L1521F-IRS, is directly detected at
mid-infrared wavelengths ($>$5 \micron) but only in scattered light at
shorter infrared wavelengths, showing a bipolar nebula oriented east-west
which is probably tracing an outflow cavity.  The nebula strongly suggests
that L1521F-IRS is embedded in the L1521F core.  Thus L1521F-IRS is similar
to the recently discovered L1014-IRS and the previously known IRAM 04191 in
its substellar luminosity and dense core environment.  However these
objects differ significantly in their core density, core chemistry, and
outflow properties, and some may be destined to be brown dwarfs rather than
stars.  
\end{abstract}

\keywords{ISM: individual (L1521F, L1521F-IRS) -- stars:
formation -- stars: low-mass, brown-dwarfs} 

\section{Introduction}
\label{sec-intro}

Studies of starless dense cores just before they form a protostar are
important as they provide the initial conditions for star formation.
However, identifying this stage of evolution is not easy.  The majority of
cores are starless in the sense that they do not harbor an IRAS point
source to a sensitivity of $L \sim$ 0.1\Lsun($d$/140 pc)$^2$ (Myers et~al.\
1987; Lee \& Myers 1999), but which starless cores are close to forming a
star?  Recently attempts have been made to identify a group of ``evolved''
starless cores with physical, chemical and kinematic properties (e.g., high
central densities, molecular depletion, infall asymmetry) that suggest they
are close to star formation (Crapsi et~al.\ 2005a; Kirk et~al.\ 2005).  It
is possible that some of these cores have already begun star formation,
containing central sources whose luminosity is too low to have been
detected by IRAS.  They are prime targets for sensitive mid-infrared
searches for very low luminosity objects with the Spitzer Space Telescope
(hereafter Spitzer).

Surprisingly, the first ``starless'' core observed by the Spitzer Legacy
program ``From Molecular Cores to Planet Forming Disks'', or c2d (Evans et
al.\ 2003), L1014, was found to contain an embedded source, L1014-IRS with
a very low luminosity  $L <$ 0.1 \Lsun\ (Young et~al.\ 2004; Bourke et~al.\
2005; Huard et~al.\ 2006).  Interestingly, L1014 shows very little evidence
of being an ``evolved'' core (Crapsi et al.\ 2005b).  If L1014-IRS and
similar objects have accretion rates which are significantly lower than
that expected of low mass protostars, then some are likely to be
proto-brown-dwarfs.  Thus, L1014-IRS and similar discoveries may pose a
significant challenge to our understanding of star formation.  

In this letter we report on our Spitzer observations of the object
L1521F-IRS which confirm that it is embedded within the dense core L1521F
in Taurus (140 pc) and has a very low luminosity.  L1521F-IRS is of
particular interest, as L1521F (a.k.a.\ MC27, Mizuno et~al.\ 1994; Codella
et~al.\ 1997; Lee, Myers \& Tafalla 2001) is one of the two best examples
of an evolved ``starless'' core, along with L1544 (Crapsi et~al.\ 2004,
2005a).  L1521F shows a high central density ($\sim10^6$ \cc), infall
asymmetry, molecular depletion, and enhanced deuterium fractionation
(Onishi, Mizuno \& Fukui 1999; Shinnaga et~al.\ 2004; Crapsi et~al.\ 2004).
Observations of CO 2-1 show no clear evidence for structured (i.e.,
bipolar) outflow emission, but line wings are seen in HCO$^+$ 3-2 (Onishi
et~al.\ 1999), which are spatially compact ($<30''$) and may be due to
outflow emission similar to that seen from L1014-IRS.

\section{Observations}

L1521F was observed by the Spitzer Space Telescope on 2004 September 7 and
9 with the Infrared Array Camera (IRAC; AORKEYs
\dataset[ADS/Sa.Spitzer#0005075712]{5075712} and 5076224), and September 24
and 25 with the Multiband Imaging Photometer (MIPS; AORKEYs 9419776 and
9431040) as part of the c2d Legacy program (Evans et~al.\ 2003).  The
observations at two closely separated epochs allow for asteroid
identification.  
The field-of-view covered by the IRAC observations was $\sim 5\farcm4
\times 10\farcm0$ at P.A.  $-8\fdg1$.  At each position four dithers of 12
s each were taken, resulting in a 48 s exposure time for most of the area
mapped.  With MIPS at 24 $\um$ a field of $\sim 9\arcmin \times 18\arcmin$
at P.A. $-10\fdg4$ was observed with a total exposure time of 48 s, and at
70 $\um$ a field of $\sim 4\farcm4 \times 12\farcm9$ at P.A. $-6\fdg9$ with
a total exposure time of 126 s was observed.  The map center for the two
epochs of MIPS observations was shifted to allow for uniform map coverage
at 70 $\um$ as half of the array is not usable.  The data were processed by
the Spitzer Science Center using their standard pipeline (version S11) to
produce Basic Calibrated Data images.  The c2d team further processed the
images to improve their quality (correcting ``bad" pixels and other array
artifacts), and performed photometry on extracted sources.  Details of the
processing, source extraction and photometry by the c2d team can be found
elsewhere (Young et al.\ 2005; Harvey et al.\ 2006).

\section{Results}

The Spitzer images of L1521F are presented in Figs.\ 1 and 2.  Figure 1
shows the 24 \micron\ and 4.5 \micron\ images of the central region of
L1521F.  At 24 \micron\ (and 70 \micron) a point source is seen located at
the dust emission peak, but it is not clearly detected above the nebular
emission at shorter wavelengths ($<$ 5 \micron).  The cross in Fig.\ 1
locates the mean position of this source, SSTc2d J042838.95+265135.1 (J2000
HHMMSS.ss+DDMMSS.s), from PSF fitting at the wavelengths at which it was
detected.  Hereafter we call this source L1521F-IRS.  At 4.5 \micron\
L1521F-IRS lies at the apex of a bright but compact conical shaped nebula,
which opens to the west.  A similar but fainter nebula is also seen opening
to the east with L1521-IRS at its apex.  This nebula is similar in
appearance to bipolar scattered light nebulae seen around young low mass
stars and also the low luminosity protostar L1014-IRS (Huard et~al.\ 2006).  

Figure 2 shows a three color image with the 3.6 \micron, 4.5 \micron, and
8.0 \micron\ bands represented by blue, green and red channels,
respectively.  Emission from all three bands is evident near the position
of L1521F-IRS (as indicated by the white emission), but the extended
emission is most evident at the shorter wavelengths (3.6-4.5 \micron).  To
the west of L1521F-IRS the emission is due to both 3.6 and 4.5 \micron\
emission (blue-green), while in the east is it mostly due to 4.5 \micron\
emission (green).

The spectral energy distribution (SED) of L1521F is shown in Figure 3, and
the flux densities are given in Table~\ref{tbl-photom}.  The measured data
points are shown as filled squares with error bars.  The 3.6 and 4.5
\micron\ points are mostly due to scattered light and so do not accurately
represent the true source flux.  The data points at 5.8 and 8.0 \micron\
also have a contribution due to scattered light, but the point-like
emission from L1521F-IRS is clearly seen on the images and its flux
dominates the emission.  The spectral index between 2 and 24 \micron\ is
estimated to be 1.6, from a linear fit to the available data points.
Although this fit is poor ($\chi^2 \sim 75$), the SED is clearly rising to
24 \micron\ and L1521F-IRS is embedded, and is consistent with a Class 0/I
object.

\section{Modeling}

The internal luminosity $L_{int}$ is a crucial property of the central
source, but cannot be obtained by simply integrating the spectrum because
the substantial contribution from the interstellar radiation field (ISRF)
must be separated out.  For this purpose a 1-D radiative transfer model is
sufficient because the envelope can be modeled as spherically symmetric
even though the disk and outflow cannot.  A detailed comparison of 1-D and
2-D models for this purpose has been undertaken by Dunham et al (2006),
indicating that only small differences in the derived values of $L_{int}$
are expected.  Consequently, we have modeled the data with the 1-D
radiative transfer code DUSTY (Ivezi\'c, Nenkova \& Elitzur 1999), in order
to attempt to place contraints on the luminosity of the central source
(star+disk), and to serve as a guide for more detailed 2-D modeling in the
future.  A full explanation of the model parameters is given by Young \&
Evans (2005), and here we briefly describe the specifics of the model used
for L1521F-IRS.  A similar analysis was performed previously for L1014-IRS
(Young et al.\ 2004) and more recently for IRAM 04191+1522 (Dunham et~al.\
2006).

The envelope is modeled using the analytic form of the density profile
given by Tafalla et al.\ (2002), $n(r)=n_0/[1+(r/r_0)^{\alpha}]$, where $n$
is the density, $r$ is the distance (radius) from the center, and $n_0,
r_0$, and $\alpha$ are free parameters.  Crapsi et al.\ (2004) fitted the
1.2 mm profile with $n_0 = 10^6$ \cc, $r_0 = 2800$ AU, and $\alpha = 2$.
The envelope is heated from the outside by the ISRF (adopting the
Black-Draine model; see Evans et al.\ 2001), and internally by a protostar
with a disk.  Because the L1521F dense core is embedded within the Taurus
molecular cloud, we attenuate the ISRF with $A_V = 3$.  We assume ``OH5''
dust opacities which are appropriate for cold, dense cores (Ossenkopf \&
Henning 1994; Shirley et al.\ 2005).  We use envelope inner and outer radii
of 10 AU (= the disk outer radius) and 14,000 AU respectively.  In order to
fit the observed long wavelength (submillimeter) data points, a value of
$n_0$ lower than that used by Crapsi et al.\ (2004) is needed, with $n_0 =
7 \times 10^5$ \cc.  With this value we obtain an envelope mass of 4.8
\Msun\ (cf., 5.5 \Msun\ in Crapsi et al.\ (2004), who used a different
value for the dust opacity).

The disk model follows the formula of Butner, Natta \& Evans (1994), with
surface density $\Sigma(r) \propto r^{-1.5}$ and temperature $T(r) \propto
r^{-0.5}$ (Beckwith et al. 1990).  The disk emission is integrated over all
possible viewing angles (i.e., a ``spherical'' disk) to compute the
emergent spectrum.  It is difficult to separate the star+disk luminosities
($L_{*} + L_{disk}$) in this model, so we treat them as a single entity,
and we consider only the luminosity of the central source $L_{int}$ ($=
L_{*} + L_{disk}$).  For the temperature of the star, $T_*$, we tried
values of 1500 and 3000 K, i.e., similar to values found for L and T
dwarfs, and late M dwarfs.  Our model does not included scattering, but the
short wavelength emission, particular at 3.6 and 4.5 \micron, is due mostly
to scattered light.  It is possible that L1521F-IRS is an almost edge-on
disk, in which case the observed short wavelength fluxes would be lower
limits.  Our model is 1-D and is an average over all inclinations, so to
allow for the edge-on possibility and obtain a strong upper limit to the
luminosity, we simply require that the model fluxes are greater than the
measured values in that wavelength range.  This requirement rules out
models with significantly lower or higher values of $T_*$, as they bring
the model fluxes uncomfortably close to the observed values.  

Our best fit models for both 1500 K and 3000 K have $L_{int} \sim 0.05
\Lsun$.  The best fit model for 1500 K is shown in Figure 3.  These models
provide reasonable matches to all wavelengths $>10$ \micron, within the
constraint required for the short wavelength emission.   At 1500 K the
maximum allowed value for $L_{int}$ is $\sim0.07$ \Lsun, which matches the
70 \micron\ flux within its $1\sigma$ uncertainty.  The best fit for 3000 K
is already close to the maximum allowed, due to the IRAC constraints -- the
maximum is $L_{int} \sim 0.06 \Lsun$.  In summary, we find that a central
source with a luminosity $L_{int} \sim 0.05 \Lsun$ provides a reasonable
fit to the data, with an upper limit of $L_{int} < 0.07 \Lsun$.  Note that
the bolometric luminosity is $L_{bol} = 0.36 \Lsun$, so that most of the
luminosity is modeled as being due to the ISRF heating the envelope (Evans
et al.\ 2001).  The short wavelength emission is not well fitted by this
simple 1-D model, which is not surprising as it is due mainly to scattered
light in the outflow cavity.  A 2-D model is required to explain the
scattered light images.  However, the main result that the luminosity of
L1521F-IRS is very low should remain unchanged in such a model (e.g.,
Dunham et al.\ 2006).

\section{Discussion}

With its low luminosity L1521F-IRS belongs to the newly identified group of
very low luminosity objects, dubbed VeLLOs, i.e., objects with $L_{int} <
0.1 \Lsun$ associated with dense cores (Kauffmann et al.\ 2005; Dunham et
al.\ 2006).  The prototype of this group is L1014-IRS with $L_{int} \sim
0.09 \Lsun$ (Young et~al.\ 2004).  The previously known object IRAM
04191+1522 (hereafter, IRAM 04191; Andr\'e, Motte \& Bacmann 1999) has
recently been shown to have a similar low luminosity, with $L_{int} \sim
0.08 \Lsun$ (Dunham et~al.\ 2006).  Although all three sources have similar
luminosities, there are a number of differences, such as their outflow
properties, which may hint at differences in their current evolutionary
status, their possible future evolution, or both.  Although current
evolutionary models of low mass objects at ages $10^5 - 10^6$ yr suggest
that all three objects are proto-brown dwarfs, the models are uncertain and
do not fully account for on-going accretion (e.g., Baraffe et al.\ 2002).
Given that all three objects possess substantial envelopes of masses a few
\Msun, their final mass could be either sub-stellar or stellar, depending
on their future accretion.

The outflow properties of these three objects deserve comparison.  IRAM
04191 has an extended molecular outflow which appears to be similar to
outflows from more luminous protostars (Andr\'e et al.\ 1999; Dunham et
al.\ 2006), while the very compact L1014-IRS outflow is less massive and
energetic than expected for its protostellar luminosity (Bourke et al.\
2005).  These results imply that the time-averaged accretion rate of IRAM
04191 ($\sim10^{-6}$ \Msun\ yr$^{-1}$; Andr\'e et al.\ 1999) is about two
orders of magnitude greater than that of L1014-IRS ($\sim10^{-8}$ \Msun\
yr$^{-1}$).  For IRAM 04191 the inferred accretion luminosity is much
greater than its internal luminosity, which may indicate the current
accretion rate is much lower (Dunham et al.\ 2006).

Although no clear bipolar molecular outflow has been detected from
L1521F-IRS, it shows a scattered light nebula which is similar in
appearance to those seen around low mass stars with outflows, and strongly
suggests that an outflow is present.  The lack of a clear outflow detection
implies that like L1014-IRS the outflow is likely to be weak and compact,
and that the accretion rate is low.  As discussed by Dunham et al.\ (2006),
the low accretion rate for L1014-IRS may hint that it is near to the end of
its main accretion phase and will remain substellar.  This could also be
the case for L1521F-IRS.  Alternatively, the present low accretion rate of
all three objects might instead indicate that they are in a quiescent phase
between episodic outbursts.

L1521F-IRS seems to be intermediate between IRAM 04191 and L1014-IRS.  The
L1521F core is bright in molecular lines and submillimeter dust emission,
and shows infall, depletion, and deuteration, (Crapsi et~al.\ 2004), all
properties seen in IRAM 04191.  However, its outflow is likely to be more
similar to the L1014-IRS outflow, and although the submillimeter core
is bright, it is not as centrally peaked as other protostellar cores, and
has been well fitted with a Bonner-Ebert-like sphere typical of starless cores
rather than a power-law typical of protostellar cores (Crapsi et al.\ 2004).  

As noted in \S\ref{sec-intro}, prior to Spitzer L1521F was one of the two
best examples of an evolved starless core, the other being the well studied
core L1544 (Tafalla et~al.\ 1998; Caselli et~al.\ 2002; Crapsi et~al.\
2005a).  Inspection of the Spitzer data for L1544 does not reveal any
protostellar candidate, i.e., L1544 is still a prestellar core.  A similar
result is found for the other evolved starless cores identified by Crapsi
et al.\ (2005).  The similarity between the L1521F and L1544 cores
suggest that L1521F-IRS has not yet had a strong influence on its parental
core, and that L1544 is indeed close to forming a protostar.   

\section{Summary \& Conclusions}

With the Spitzer Space Telescope we have discovered a low luminosity object
L1521F-IRS ($L_{int} = 0.05 \Lsun$, with an upper limit of $\sim 0.07$
\Lsun) within the ``evolved starless core'' L1521F in Taurus.  At short
wavelengths ($<$ 5 \micron) a well defined bipolar scattered light nebula
is seen, suggesting the presence of a molecular outflow like those observed
from low mass protostars.  L1521F-IRS is located at the density peak of the
core, and its luminosity is similar to other low luminosity objects IRAM
04191+1522 and L1014-IRS.  It is unclear at present whether L1521F-IRS will
evolve to form a low mass star or a brown dwarf.

These results do not support the idea that core properties correlate with
features of early stellar evolution.  Some cores, including L1544, have
hallmarks of high density, including freeze-out chemistry, but harbor no
embedded sources, while similar cores do have internal sources, such as
L1521F. Other cores, including L1014, have modest densities yet harbor
VeLLOs.  Still other cores, such as L1521F and IRAM 04191, have
similar gas properties and objects of similar luminosity, but their
embedded sources have very different mass outflow rates.  It is not clear
where VeLLOs fit within the current paradigm of low
mass star formation, and it is difficult to imagine a single evolutionary
scenario which can explain them all.  The picture may become clearer
once a large enough sample of these objects has been studied.

\acknowledgments

Support for this work, part of the Spitzer Legacy Science Program, was
provided by NASA through contracts 1224608, 1230782, and 1230779 issued by
the Jet Propulsion Laboratory, California Institute of Technology, under
NASA contract 1407.  Support from NASA Origins Grants NAG5-13050 (P.C.M.)
and NNG04GG24G (N.J.E.) are also acknowledged.  C.W.L acknowledges KOSEF
grant R01-2003-000-10513-0.  We thank the Lorentz Center in Leiden for
hosting several meetings that contributed to this paper, and Rob Gutermuth
for supplying the IDL code used to make the color figure.


\begin{deluxetable}{lrrc}
\tablewidth{3.25in}
\tablecaption{Photometry of L1521F\label{tbl-photom}}
\tablehead{
\colhead{Wavelength} & \colhead{Flux Density} & 
\colhead{$\sigma$} & \colhead{Aperture} \\
\colhead{(\micron)} & \colhead{(mJy)} & \colhead{(mJy)} &
\colhead{(arcsec)}
}
\startdata
3.6         & 0.42      & 0.06		& 1.7\tablenotemark{a} \\
4.5         & 0.54	& 0.08		& 1.7\tablenotemark{a} \\
5.8         & 0.34	& 0.05		& 1.9\tablenotemark{a} \\
8.0         & 0.45	& 0.07		& 2.0\tablenotemark{a} \\
24          & 25.0	& 4		& 5.7\tablenotemark{a} \\
70          & 490	& 200		& 17\tablenotemark{a}  \\
350	    & 4600	& 700 		& 40 \\
450         & 7000      & 1800          & 40 \\
850         & 1500	& 500		& 40 \\
1200        & 600	& 150		& 40 \\
%
\enddata
\tablenotetext{a}{FWHM of the Spitzer point spread profile.}
\tablecomments{(1) The listed uncertainties include the statistical measurement
errors and uncertainties in the absolute calibration.
(2) The (sub)mm points were obtained with the bolometer array cameras
SHARCII (350 \micron\ -- J.\ Wu et~al.\ in prep.), SCUBA (850 \micron\ --
C.\ Young et~al.\ in prep.; 450 \micron -- Kirk, Ward-Thompson \& Andr\'e
2005; J.\ Kirk, private communication) and MAMBO (1200 \micron\ -- Crapsi
et~al.\ 2004; J.\ Kauffmann et al.\ in prep.).  
}
\end{deluxetable}

\begin{figure}[!t]
\centering
\includegraphics[width=3.25in]{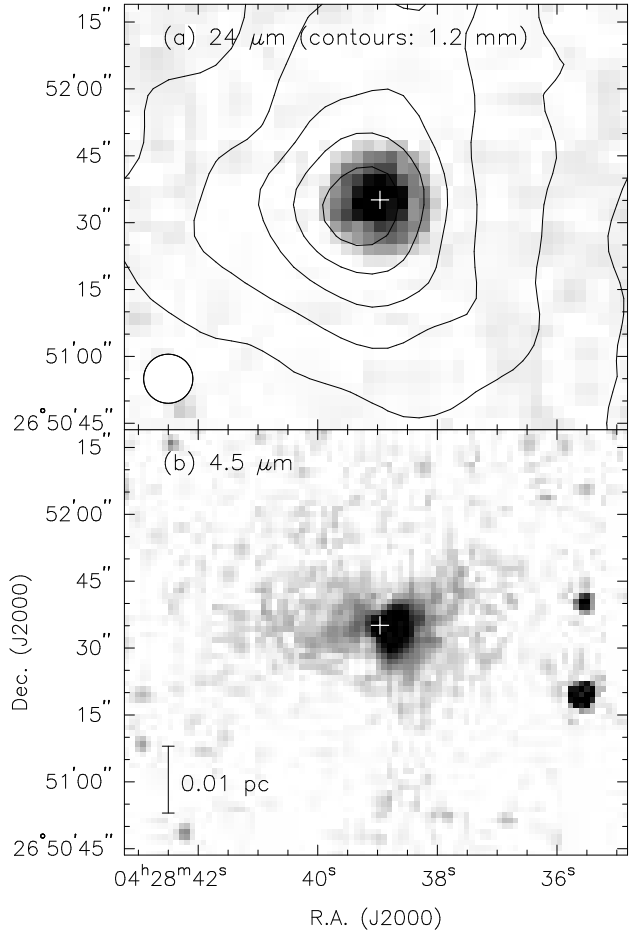}
\caption{
(a) MIPS 1 (24 \micron) image overlain with contours of 1.2 mm emission
(Crapsi et~al.\ 2004; J.~Kauffmann et~al.\ in prep).  
The contour levels are 15,30,...,90\% of the peak of 94 mJy beam$^{-1}$.
The white cross indicates position of L1521F-IRS in both panels, and the
1.2 mm beam is shown as the circle at lower left.  (b) IRAC 2 (4.5 \um)
image.  
\label{fig-mips}}
\end{figure}

\begin{figure}[t]
\centering
\includegraphics[width=3.25in]{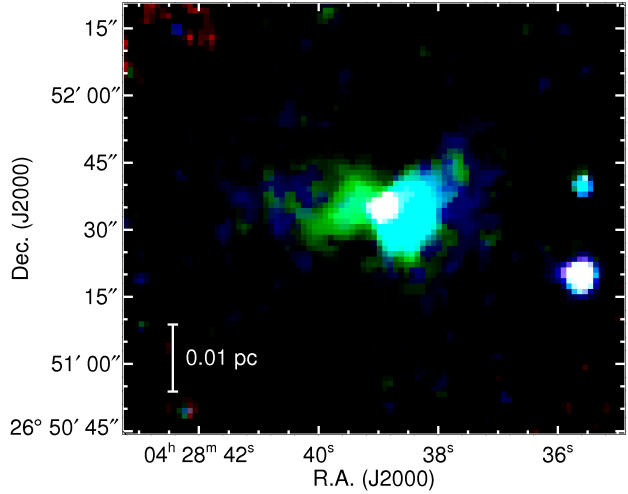}
\caption{
Three color image of the central region of L1521F, using IRAC 1 (3.6 \um;
blue), IRAC 2 (4.5 \um; green) and IRAC 4 (8.0 um; red).
\label{fig-rgb}}
\end{figure}

\begin{figure}[!t] 
\centering
\includegraphics[height=3.25in,angle=270]{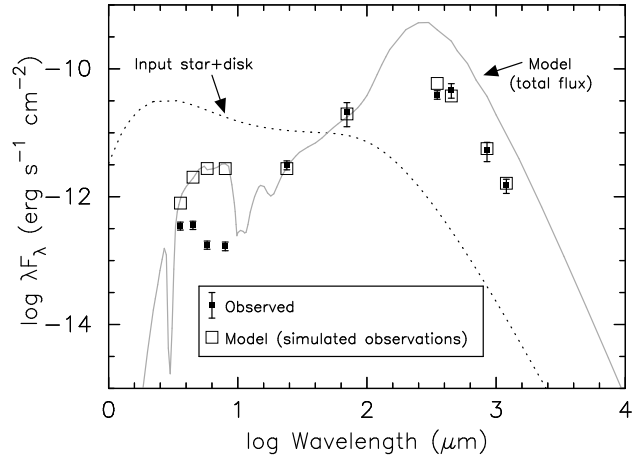}
\caption{
SED of L1521F-IRS with a 1-D model fit.  The solid squares with error bars
are the measured flux densities from Table~\ref{tbl-photom}.  At (sub)mm
wavelengths the fluxes within 40\arcsec\ apertures are shown.  The dotted
line is the input star+disk spectrum, and the total flux density of the
output model spectrum is shown by the faint grey line.  For comparison with
the observed fluxes, the model fluxes were convolved with the same
photometric filters (Spitzer) or observed with the same apertures (submm),
and are shown as open squares.  
%
\label{fig-sed}}
\end{figure}

\end{document}